\documentclass[twocolumn,aps,pra,amsmath,amssymb,notitlepage]{revtex4-1}
\usepackage[latin9]{inputenc}
\usepackage{amsmath}
\usepackage{amssymb}
\usepackage{graphicx}
\usepackage{wasysym}
\usepackage{esint}

\makeatletter

\newcommand*\LyXThinSpace{\,\hspace{0pt}}

\usepackage{mathrsfs}

\usepackage{epsfig}
\usepackage{bm}
\usepackage{dcolumn}
\usepackage{color}

\makeatother

\begin{document}
\title{Ultradilute self-bound quantum droplets in Bose-Bose mixtures at finite
temperature}
\author{Jia Wang, Xia-Ji Liu, and Hui Hu}
\affiliation{Centre for Quantum Technology Theory, Swinburne University of Technology,
Melbourne, Victoria 3122, Australia}
\date{\today}
\begin{abstract}
We theoretically investigate the finite-temperature structure and
collective excitations of a self-bound ultradilute Bose droplet in
a flat space realized in a binary Bose mixture with attractive inter-species
interactions on the verge of mean-field collapse. As the droplet formation
relies critically on the repulsive force provided by Lee-Huang-Yang
quantum fluctuations, which can be easily compensated by thermal fluctuations,
we find a significant temperature effect in the density distribution
and collective excitation spectrum of the Bose droplet. A finite-temperature
phase diagram as a function of the number of particles is determined.
We show that the critical number of particles at the droplet-to-gas
transition increases dramatically with increasing temperature. Towards
the bulk threshold temperature for thermally destabilizing an infinitely
large droplet, we find that the excitation-forbidden, self-evaporation
region in the excitation spectrum, predicted earlier by Petrov using
a zero-temperature theory, shrinks and eventually disappears. All
the collective excitations, including both surface modes and compressional
bulk modes, become softened at the droplet-to-gas transition. The
predicted temperature effects of a self-bound Bose droplet in this
work could be difficult to measure experimentally due to the lack
of efficient thermometry at low temperatures. However, these effects
may already present in the current cold-atom experiments. 
\end{abstract}
\maketitle

\section{Introduction}

The recent observation of an ultradilute self-bound droplet-like state
\cite{Bottcher2020} in single-component dipolar Bose-Einstein condensates
(BECs) \cite{FerrierBarbut2016,Schmitt2016,Chomaz2016,Bottcher2019}
and binary Bose-Bose mixtures \cite{Cabrera2018,Cheiney2018,Semeghini2018,Ferioli2019,DErrico2019,Wang2019IASWorkshop}
opens an entirely new direction to better understand the fascinating
concept of quantum droplets - autonomously isolated quantum systems
equilibrated under \emph{zero} pressure in free space. Quantum droplets
such as helium nano-droplets have already been intensively investigated
in condensed matter community over the past few decades \cite{Barranco2006,Gessner2019}.
However, an in-depth understanding of helium nano-droplets is still
lacking, due to the strong inter-particle interactions and the limited
techniques to control and characterize the nano-droplets. These limitations
could be overcome for ultradilute Bose droplets, owing to the unprecedented
controllability in cold-atom experiments \cite{Dalfovo1999}. For
example, the inter-particle interactions in Bose droplets can be tuned
at will by using Feshbach resonances \cite{Chin2010} and their structure
and collective excitations can be accurately measured through in-situ
or time-of-flight absorption imaging \cite{Dalfovo1999}. In particular,
the realization of a weakly interacting Bose droplet now allows us
to develop quantitative descriptions and make it possible to have
testable theoretical predictions \cite{Petrov2015,Petrov2016,Baillie2016,Wachtler2016,Li2017,Cappellaro2018,Astrakharchik2018,Cui2018,Staudinger2018,Ancilotto2018,Parisi2019,Aybar2019,Cikojevic2019,Chiquillo2019,Minardi2019,Tylutki2020,Hu2020a,Hu2020b,Hu2020c,Wang2020PRR}.

In this respect, it is worth noting the seminal work by Petrov \cite{Petrov2015},
where the existence of a Bose droplet is proposed in binary Bose mixtures
with attractive inter-species attractions. The mean-field collapse
is surprisingly shown to be arrested by an effective repulsive force
arising from Lee-Huang-Yang (LHY) quantum fluctuations \cite{LeeHuangYang1957}.
This ground-breaking proposal is now successfully confirmed in several
experimental setups, including the homonuclear $^{39}$K-$^{39}$K
mixtures \cite{Cabrera2018,Semeghini2018,Cheiney2018,Ferioli2019}
and heteronuclear $^{41}$K-$^{87}$Rb \cite{DErrico2019} or $^{23}$Na-$^{87}$Rb
mixtures \cite{Wang2019IASWorkshop}. Following Petrov's pioneering
idea \cite{Petrov2015}, numerous theoretical investigations have
been recently carried out \cite{Petrov2015,Petrov2016,Baillie2016,Wachtler2016,Li2017,Cappellaro2018,Astrakharchik2018,Cui2018,Staudinger2018,Ancilotto2018,Parisi2019,Cikojevic2019,Chiquillo2019,Minardi2019,Tylutki2020,Hu2020a,Hu2020b,Hu2020c,Wang2020PRR},
addressing various zero-temperature properties of Bose droplets.

The finite-temperature properties of Bose droplets in both dipolar
BECs and binary Bose mixtures, however, do not receive too much attention.
Ultradilute droplets of dipolar bosons at finite temperature have
recently been considered in the presence of an external harmonic trap
\cite{Aybar2019}. For Bose droplets in binary mixtures, only the
\emph{bulk} properties (of an infinitely large droplet) at nonzero
temperature are addressed most recently \cite{Wang2020,Ota2020}.
As the effective repulsive force provided by the LHY fluctuation term
can be easily neutralized by thermal fluctuations, it is not a surprise
to find that a Bose droplet in binary mixtures can be completely destabilized
above a threshold temperature $T_{\textrm{th}}$ \cite{Wang2020}.

The less interest in the finite temperature effect is probably due
to the peculiar \emph{self-evaporation} feature of quantum droplets.
As a self-bound entity, the energy of elementary excitations of quantum
droplets - either single-particle excitations or collective excitations
- has to be bounded from above by the so-called particle-emission
threshold, making the droplet essentially a low-temperature object.
This is fairly evident in helium nano-droplets: once the nano-droplet
is created, its temperature rapidly decreases to about $0.4$ Kelvin
in several milliseconds \cite{Barranco2006}. Afterward, however,
the self-evaporation becomes not so efficient \cite{Barranco2006}.
A Bose droplet in a binary mixture is similarly anticipated to be
a low-temperature object. In particular, as predicted by Petrov from
zero-temperature calculations \cite{Petrov2015}, for the number of
particles in a certain range, there are no collective excitations
below the particle-emission threshold. In other words, an excitation-forbidden
region in the particle number exists. The Bose droplet then may automatically
lose its thermal energy upon releasing the most energetic particles
and reach exactly zero temperature.

In this work, we would like to argue that the self-evaporation efficiency
of Bose droplets at low temperature could be much reduced, as in helium
nano-droplets \cite{Barranco2006}. As a result, the experimentally
observed Bose droplets might have a small but nonzero temperature
in the realistic time-scale of experiments \cite{Cabrera2018,Semeghini2018,Cheiney2018,Ferioli2019}.
We theoretically determine the finite-temperature structure and collective
excitations of self-bound spherical Bose droplets with a finite number
of particles, based on time-independent and time-dependent extended
Gross-Pitaevskii equations (GPEs) \cite{Petrov2015}, respectively.
We find a rich phase diagram at finite temperature. In particular,
the excitation-forbidden, self-evaporation region of the Bose droplet,
found earlier by Petrov using a zero-temperature theory \cite{Petrov2015},
turns out to shrink with increasing temperature and disappears eventually.
We also predict that the surface modes and compressional sound modes
of the Bose droplet become softened at the droplet-to-gas transition
upon increasing temperature. Our results could be experimentally examined
in binary Bose mixtures if efficient thermometry can be established
at low temperatures.

\section{Extended Gross-Pitaevskii equation at finite temperature}

To address the finite-temperature properties of a finite-size Bose
droplet, we consider the \emph{finite-temperature} version of the
extended GPE, 
\begin{equation}
i\hbar\frac{\partial\Phi}{\partial t}=\left[-\frac{\hbar^{2}}{2m}\nabla^{2}-\mu+\frac{\partial\mathcal{F}}{\partial n}\left(n=\left|\Phi\right|^{2}\right)\right]\Phi,\label{eq:EGPE}
\end{equation}
where $\Phi(\mathbf{x},t)$ can be treated as the wave-function of
the Bose droplet with atomic mass $m$, $\mu$ is the chemical potential
to be determined by the total number of particles $N$, and $\mathcal{F}(n)$
is the local \emph{free} energy functional (per unit volume $\mathcal{V}$)
depending on the local density $n(\mathbf{x},t)=\left|\Phi(\mathbf{x},t)\right|^{2}$.
At zero temperature, the free energy functional $\mathcal{F}(n)$
reduces to the ground-state energy functional $\mathcal{E}(n)$ \cite{Petrov2015},
and we recover the zero-temperature extended GPE \cite{Petrov2015}
that has been extensively used in the literature.

\begin{figure}[t]
\begin{centering}
\includegraphics[width=0.48\textwidth]{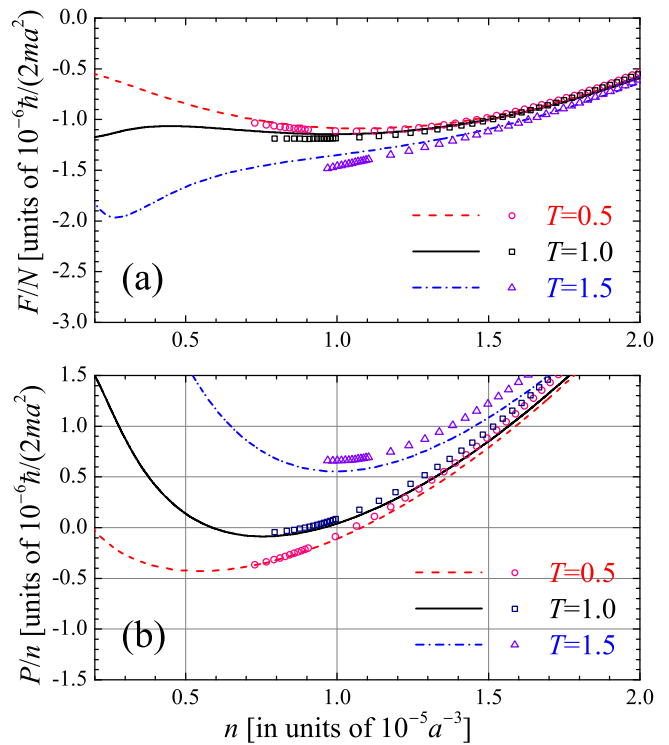} 
\par\end{centering}
\caption{\label{fig1_eosfitting} Equation of state (EoS). The free energy
per particle $\mathcal{F}/n$ (a) and pressure per particle $P/n$
(b) of a homogeneous binary Bose mixture, in units of $10^{-6}\hbar^{2}/(2ma^{2})$,
are shown as a function of the density $n$ at three temperatures
$k_{B}T=0.5$, $1.0$ and $1.5$. The density $n$ and temperature
$T$ are measured in units of $10^{-5}a^{-3}$ and $10^{-4}\hbar^{2}/(2ma^{2})$,
respectively. A self-bound Bose droplet is realized, when the pressure
is zero at an equilibrium density $n_{0}$. The symbols and lines
show the numerical and analytical results, respectively. The former
is from a bosonic pairing theory, while the latter is calculated according
to Eq. (\ref{eq:FreeEnergyFunctional}) and Eq. (\ref{eq:s3}). We
note that the numerical results are not available at sufficiently
low density. Here, we consider the inter-species interaction strength
$a_{12}=-1.05a$. }
\end{figure}

In a binary Bose mixture, the extended GPE Eq. (\ref{eq:EGPE}) for
Bose droplets can be microscopically derived by using a bosonic pairing
theory \cite{Hu2020a,Hu2020b}. This was demonstrated in the recent
work at zero temperature under the local density approximation \cite{Hu2020c}.
The generalization of such a derivation to the nonzero temperature
$T\neq0$ is straightforward. It is easy to show that the free energy
functional per unit volume could take the form \cite{Wang2020}, 
\begin{eqnarray}
\mathcal{F} & = & -\frac{\pi\hbar^{2}}{m}\left(a+\frac{a^{2}}{a_{12}}\right)n^{2}+\frac{256\sqrt{\pi}}{15}\left(\frac{\hbar^{2}a^{5/2}}{m}\right)n^{5/2}\nonumber \\
 &  & -\frac{\sqrt{\pi}}{720}\left(\frac{m^{3}k_{B}^{4}T^{4}}{\hbar^{6}a^{3/2}}\right)n^{-3/2}\tilde{s}_{3}\left(1,\gamma\right),\label{eq:FreeEnergyFunctional}
\end{eqnarray}
where $a>0$ and $a_{12}\simeq-a$ are the intra-species and inter-species
$s$-wave scattering lengths of the binary Bose mixture, respectively.
The last term in the above expression accounts for the finite-temperature
effect and the function $\tilde{s}_{3}[1,\gamma\equiv mk_{B}T/(2\pi\hbar^{2}an)]$
could be calculated numerically \cite{Wang2020}. It becomes unity
in the zero-temperature limit and decreases with increasing temperature
\cite{Wang2020}. It turns out that close to the threshold for the
formation of Bose droplets, i.e., $a_{12}\simeq-a$, the function
$\tilde{s}_{3}(1,\gamma)$ could be reasonably approximated by an
exponential decay form at low temperature, i.e., 
\begin{equation}
s_{3}\left(1,\gamma\right)\simeq\exp\left(-A\gamma\right),\label{eq:s3}
\end{equation}
with a decay constant $A\simeq0.5$. This is shown in Fig. \ref{fig1_eosfitting}(a),
where the predictions of Eq. (\ref{eq:FreeEnergyFunctional}) and
Eq. (\ref{eq:s3}) are compared with the numerical results from the
bosonic pairing theory \cite{Wang2020}. Eq. (2) also enables us to
calculate the pressure by using the standard thermodynamic relation,
$P=n^{2}[\partial(\mathcal{F}/n)/\partial n]$. It is evident from
Fig. \ref{fig1_eosfitting}(b) that the zero-pressure condition, which
is required for a self-bound Bose droplet, can not be satisfied for
relatively large temperature (i.e., $T=1.5\times10^{-4}\hbar^{2}/(2ma^{2})$).
Indeed, from the analysis of our previous work \cite{Wang2020}, we
find that an infinitely large Bose droplet thermally dissolves itself
once the temperature is above a threshold, 
\begin{equation}
k_{B}T_{\textrm{th}}\simeq0.0222\left(1+\frac{a}{a_{12}}\right)^{2}\frac{\hbar^{2}}{ma^{2}}.\label{eq:TTC3D}
\end{equation}
For the inter-species interaction strength considered in Fig. \ref{fig1_eosfitting},
$a_{12}=-1.05a$, the threshold temperature is about $k_{B}T_{\textrm{th}}\simeq5.03\times10^{-5}\hbar^{2}/(ma^{2})$.
To be specific, we will concentrate on this example with $a_{12}=-1.05a$.
But, it will become clear later that our results do not depend on
this particular choice of the interaction parameters.

It is useful to note that an exponential decay form of the function
$\tilde{s}_{3}(1,\gamma)$ is very helpful for our numerical solutions
of the extended GPE. This exponential decay, i.e., $e^{-B/n}$ with
$B\equiv Amk_{B}T/(2\pi\hbar^{2}a)$, largely compensates the power-law
increase $n^{-3/2}$ in the last term of the free energy Eq. (\ref{eq:FreeEnergyFunctional})
at low density and therefore remove the possible numerical instability
caused by the sufficiently low density at the edge of the Bose droplet.

To ease the numerical workload, it is also helpful to use the (dimensionless)
re-scaled coordinate and time as suggested by Petrov in his pioneering
work \cite{Petrov2015}: 
\begin{eqnarray}
\tilde{\mathbf{x}} & = & \frac{\mathbf{x}}{\xi},\\
\tilde{t} & = & \frac{t}{m\xi^{2}/\hbar},\\
\phi & = & \frac{\Phi}{\sqrt{n_{0}}},\\
\tilde{\mu} & = & \frac{\mu}{\hbar^{2}/\left(m\xi^{2}\right)},
\end{eqnarray}
where the equilibrium (bulk) density $n_{0}$ of a large Bose droplet
at zero temperature is \cite{Hu2020a} 
\begin{equation}
n_{0}\equiv\frac{25\pi}{16384}\left(1+\frac{a}{a_{12}}\right)^{2}a^{-3}\label{eq:n0}
\end{equation}
and the length scale $\xi$ can be chosen in such a way that the ground-state
energy functional (i.e., the free energy functional at zero temperature)
per unit volume takes the form $\mathcal{E}=-3\left|\phi\right|^{4}/2+\left|\phi\right|^{5}$
\cite{Petrov2015}. This leads to the energy scale, 
\begin{equation}
\frac{\hbar^{2}}{m\xi^{2}}=\frac{25\pi^{2}}{24576}\left(1+\frac{a}{a_{12}}\right)^{3}\frac{\hbar^{2}}{ma^{2}}.\label{eq:EnergyScale}
\end{equation}
At $a_{12}=-1.05a$, we find the energy units $\hbar^{2}/(m\xi^{2})\simeq1.08\times10^{-6}\hbar^{2}/(ma^{2})$
and the units for the number of particles $n_{0}\xi^{3}=N/\tilde{N}\simeq9630$,
where $\tilde{N}$ is the reduced number of particles. In terms of
the energy scale $\hbar^{2}/(m\xi^{2})$, we may re-write the threshold
temperature, 
\begin{equation}
k_{B}T_{\textrm{th}}\simeq2.21\left(1+\frac{a}{a_{12}}\right)^{-1}\frac{\hbar^{2}}{m\xi^{2}},\label{eq:TemperatureScale}
\end{equation}
which is about $46.4\hbar^{2}/(m\xi^{2})$ at $a_{12}=-1.05a$.

From now on, without any confusion we shall remove the tilde above
the re-scaled quantities. The time-dependent extended GPE then takes
the following \emph{dimensionless} form,

\begin{equation}
i\frac{\partial}{\partial t}\phi=\left[-\frac{1}{2}\nabla^{2}-\mu+\frac{\partial\mathcal{F}}{\partial n}\left(\phi,\phi^{*}\right)\right]\phi,\label{eq:dimensionlessEGPE}
\end{equation}
where the dimensionless free energy functional is now given by (the
density $n=\left|\phi\right|^{2}$), 
\begin{equation}
\mathcal{F}=-\frac{3}{2}\left|\phi\right|^{4}+\left|\phi\right|^{5}-\frac{C_{T}}{\left|\phi\right|^{3}}\exp\left[-\frac{B_{T}}{\left|\phi\right|^{2}}\right]\label{eq:DimensionlessFreeEnergy}
\end{equation}
with the temperature-dependent constants, 
\begin{eqnarray}
C_{T} & \equiv & \frac{\pi^{4}}{62208}\left[\frac{\left(1+a/a_{12}\right)k_{B}T}{\hbar^{2}/\left(m\xi^{2}\right)}\right]^{4}\simeq0.0374\frac{T^{4}}{T_{\textrm{th}}^{4}},\\
B_{T} & \equiv & \frac{A}{3}\left(1+\frac{a}{a_{12}}\right)\frac{k_{B}T}{\hbar^{2}/\left(m\xi^{2}\right)}\simeq0.369\frac{T}{T_{\textrm{th}}}.
\end{eqnarray}
At zero temperature, where the last term in Eq. (\ref{eq:DimensionlessFreeEnergy})
is absent, Eq. (\ref{eq:dimensionlessEGPE}) recovers the dimensionless
extended GPE used earlier by Petrov \cite{Petrov2015}. The two temperature-dependent
constants $C_{T}$ and $B_{T}$ are the function of the ratio $T/T_{\textrm{th}}$
only and do not depend explicitly on the scattering lengths $a$ and
$a_{12}$. Near the threshold temperature $T\sim T_{\textrm{th}}$,
both constants become significant, and we find that the LHY quantum-fluctuation
term ($\propto\left|\phi\right|^{5}$) in the free energy functional
is largely compensated by the last thermal-fluctuation term. This
eventually destabilizes a large Bose droplet at the threshold temperature
$T_{\textrm{th}}$ \cite{Wang2020}. Let us now analyze how does this
thermal destabilization occur for a Bose droplet with a finite (reduced)
number of particles.

\subsection{Time-independent GPE for the density distribution $\phi_{0}$}

In free space, the self-bound Bose droplet takes an isotropic spherical
profile, depending on the radius $r$ only \cite{Semeghini2018,Petrov2015}.
The static profile $\phi_{0}(r)\geq0$ at finite temperature satisfies
the time-independent version of Eq. (\ref{eq:dimensionlessEGPE}):

\begin{equation}
\mathcal{\hat{L}}\phi_{0}\left(r\right)=\mu\phi_{0}\left(r\right),\label{eq:stationaryEGPE}
\end{equation}
where we have defined the operator, 
\begin{equation}
\mathcal{\hat{L}}\equiv-\frac{\nabla^{2}}{2}-3\phi_{0}^{2}+\frac{5}{2}\phi_{0}^{3}+\frac{\partial\mathcal{F}_{T}}{\partial n}.
\end{equation}
The last term of the thermal contribution to $\hat{\mathcal{L}}$
is explicitly given by, 
\begin{equation}
\frac{\partial\mathcal{F}_{T}}{\partial n}=+\frac{3C_{T}}{2\phi_{0}^{5}}\left(1-\frac{2B_{T}}{3\phi_{0}^{2}}\right)\exp\left[-\frac{B_{T}}{\phi_{0}^{2}}\right].
\end{equation}
In this work, we solve the static GPE numerically via a gradient method,
which improves the accuracy and efficiency from our previous work
\cite{Dalfovo1999,Pu1998,Hu2020d}. The details of our numerical method
are described in the Appendix.

\subsection{Bogoliubov equations for collective excitations}

To study the collective excitations of the Bose droplet, we consider
small fluctuation modes around the condensate wave-function $\phi_{0}(r)$
\cite{Dalfovo1999,Tylutki2020,Hu2020d}, 
\begin{equation}
\phi(\mathbf{x},t)=\phi_{0}(r)+\sum_{j}\left[u_{j}\left(\mathbf{x}\right)e^{-i\omega_{j}t}+v_{j}^{*}\left(\mathbf{x}\right)e^{+i\omega_{j}t}\right],
\end{equation}
where different mode with mode frequency $\omega_{j}$ is indexed
by an integer $j$, and $u_{j}(\mathbf{x})$ and $v_{j}(\mathbf{x})$
are the corresponding mode wave-functions. For a spherical droplet,
the index $j$ can be further denoted by two good quantum numbers
($l,n$), where $l$ is the angular momentum and $n$ stands for the
radial quantum number (i.e., the number of nodes in the radial wave-function).
By substituting the above expression into the dimensionless extended
GPE Eq. (\ref{eq:dimensionlessEGPE}) and expanding it to the linear
order in $u_{j}(\mathbf{x})$ and $v_{j}(\mathbf{x})$, we obtain
the celebrated Bogoliubov equations \cite{Dalfovo1999,Tylutki2020,Hu2020d},
\begin{equation}
\left[\begin{array}{cc}
\mathcal{\hat{L}}-\mu+\mathcal{\hat{M}} & \mathcal{\hat{M}}\\
\mathcal{\hat{M}} & \mathcal{\hat{L}}-\mu+\mathcal{\hat{M}}
\end{array}\right]\left[\begin{array}{c}
u_{j}\left(\mathbf{r}\right)\\
v_{j}\left(\mathbf{r}\right)
\end{array}\right]=\omega_{j}\left[\begin{array}{c}
+u_{j}\left(\mathbf{r}\right)\\
-v_{j}\left(\mathbf{r}\right)
\end{array}\right],\label{eq:BogoliubovEQs}
\end{equation}
where the operator $\hat{\mathcal{M}}$ is given by, 
\begin{equation}
\mathcal{\hat{M}}\equiv n\frac{\partial^{2}\mathcal{F}}{\partial n^{2}}=-3\phi_{0}^{2}+\frac{15}{4}\phi_{0}^{3}+n\frac{\partial^{2}\mathcal{F}_{T}}{\partial n^{2}},
\end{equation}
and the explicit form of the last thermal term in $\hat{\mathcal{M}}$
is, 
\begin{equation}
n\frac{\partial^{2}\mathcal{F}_{T}}{\partial n^{2}}=-\frac{15C_{T}}{4\phi_{0}^{5}}\left(1-\frac{4B_{T}}{3\phi_{0}^{2}}+\frac{4B_{T}^{2}}{15\phi_{0}^{4}}\right)\exp\left[-\frac{B_{T}}{\phi_{0}^{2}}\right].
\end{equation}
It should be noted that the wave-functions $u(\mathbf{x})=+\phi_{0}(r)$
and $v(\mathbf{x})=-\phi_{0}(r)$ are the zero-energy solution (i.e.,
$\omega_{j}=0$) of the Bogoliubov equations. This is precisely the
condensate mode of the Bose droplet and therefore should be discarded.
To numerically solve the Bogoliubov equations, we follow the technique
by Hutchinson, Zaremba and Griffin \cite{Hutchinson1997}. The details
of the numerical implementation can be found in Ref. \cite{Hu2020d}
and the Appendix.

\section{Results and discussions}

For a given reduced number of particles $N$ and a given reduced temperature
(i.e., the ratio $T/T_{\textrm{th}}$), we have numerically solved
the dimensionless static GPE Eq. (\ref{eq:stationaryEGPE}) and Bogoliubov
equations Eq. (\ref{eq:BogoliubovEQs}) for the density distribution
$\phi_{0}(r)$ and the collective excitation spectrum $\omega_{j}$,
respectively. To connect with the experimental observables at different
scattering lengths $a$ and $a_{12}$, we can restore the units of
different quantities (i.e., density, mode frequency and temperature)
by simply multiplying, for example, the equilibrium density $n_{0}$,
the energy scale $\hbar^{2}/(m\xi^{2})$ and the temperature scale
$T_{\textrm{th}}$, which are given in Eq. (\ref{eq:n0}), Eq. (\ref{eq:EnergyScale})
and Eq. (\ref{eq:TemperatureScale}), respectively.

In the following, we first consider the excitation spectrum at some
chosen temperatures and determine a rich finite-temperature phase
diagram. We then discuss in detail the temperature dependence of the
density distribution and excitation spectrum at fixed number of particles,
mimicking the realistic experimental measurements with running temperature.

\begin{figure}[t]
\begin{centering}
\includegraphics[width=0.48\textwidth]{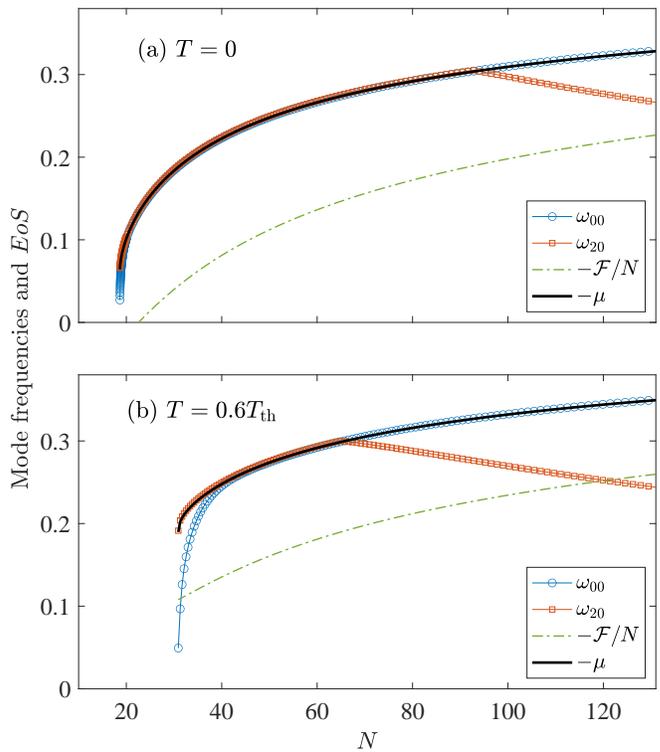} 
\par\end{centering}
\caption{\label{fig2_CMLowTemp} The chemical potential $-\mu$ (black solid
curves), free energy per particle $-\mathcal{F}/N$ (green dash-dotted
curves), breathing mode frequency (blue open circles with curves),
and the quadrupole mode frequency (red open squares with curves) as
a function of the number of particle at zero temperature (a) and at
temperature $T=0.6T_{\textrm{th}}$ (b). At zero temperature in (a),
a metastable Bose droplet occurs when the number of particles decreases
down to $N_{m}\simeq22.55$, when the free energy $\mathcal{F}$ becomes
positive (or $-\mathcal{F}$ becomes negative).}
\end{figure}

\begin{figure}[t]
\begin{centering}
\includegraphics[width=0.48\textwidth]{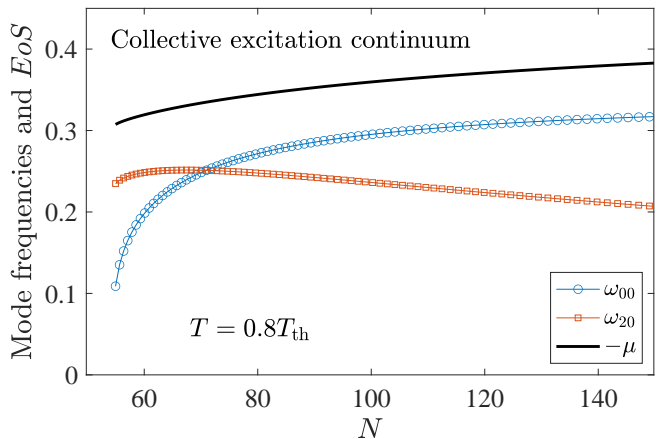} 
\par\end{centering}
\caption{\label{fig3_CMLargeTemp} The chemical potential $-\mu$ (black lines),
breathing mode frequency (blue solid circles with lines), and the
quadrupole mode frequency (red open squares with lines) as a function
of the number of particle at temperature $T=0.8T_{\textrm{th}}$ (b).}
\end{figure}

\subsection{Collective excitations at a given temperature}

To start, let us briefly review the essential zero-temperature properties
of a self-bound Bose droplet \cite{Petrov2015}. First, the chemical
potential $\mu$ of the droplet has to be negative ($\mu<0$), less
than that of the surrounding vacuum (i.e., $\mu_{\textrm{vac}}=0$).
Otherwise, it is not energetically favorable for particles to be added
into the droplet. For an infinitely large droplet, where the edge
effect can be safely neglected, it is clear from the stationary GPE
Eq. (\ref{eq:stationaryEGPE}) that the condensate wave-function in
the bulk is $\phi_{0}=1$ in the re-scale units and the chemical potential
$\mu=-1/2$. As we decrease the number of particles in the droplet,
the wave-function $\phi_{0}(r)$ will be smaller than unity and the
chemical potential increases towards $\mu\rightarrow0^{-}$. The droplet
will eventually become unstable and experience a droplet-to-gas transition,
when the zero-pressure condition for the droplet state is strongly
violated at low density. The droplet-to-gas transition at the critical
number of particles $N_{c}\simeq18.65$ has been analyzed in detail
by Petrov \cite{Petrov2015}, by considering the balance between the
kinetic energy (i.e. from the Laplace operator $-\nabla^{2}/2$) and
the interaction energy (i.e., $\mathcal{F}(\phi_{0}))$. The transition
is clearly signaled by the softening of the breathing mode frequency
$\omega_{l=0,n=0}$, which vanishes precisely at $N_{c}$. This is
shown by blue circles in Fig. \ref{fig2_CMLowTemp}(a), where we reproduce
the lower panel of Fig. 1 in Ref. \cite{Petrov2015}. Petrov also
predicted the existence of a metastable droplet state when the number
of particles is slightly larger than the critical number, i.e., $N_{c}<N<N_{m}\simeq22.55$
\cite{Petrov2015}. This metastable state has a positive total energy,
i.e., $-\mathcal{F}<0$ as shown in Fig. \ref{fig2_CMLowTemp}(a),
so the particles in the droplet will eventually escape to the vacuum
via tunneling through an energy barrier (created by the competing
kinetic and interaction energies).

Another interesting zero-temperature feature of the Bose droplet is
the existence of an excitation-forbidden window in the number of particles
\cite{Petrov2015}, as we mentioned earlier. In Fig. \ref{fig2_CMLowTemp}(a),
there is no collective excitations in the stable droplet state below
a threshold number of particles, $N<N_{\textrm{th}}\simeq94.2$. All
collective excitations are accumulated right above the particle-emission
threshold $\left|\mu\right|$, forming an \emph{unbounded} collective
excitation continuum \cite{Hu2020d}. The bounded collective excitations
are only possible at $N>N_{\textrm{th}}$, where the quadruple mode
frequency $\omega_{l=2,n=0}$ first becomes smaller than $\left|\mu\right|$
\cite{Petrov2015}, as shown by the red empty squares. At sufficiently
large number of particles (i.e., $N\sim10^{4}$), the Bose droplet
is able to acquire a series of the surface modes and compressional
sound modes with well-defined dispersion relations \cite{Petrov2015,Hu2020d},
as we shall see later.

At finite temperature, the collective excitation spectrum can dramatically
change. In Fig. \ref{fig2_CMLowTemp}(b), we report the excitation
spectrum at $T=0.6T_{\textrm{th}}$. It is readily seen that the droplet-to-gas
transition now occurs at a much larger critical number of particles,
$N_{c}(T=0.6T_{\textrm{th}})\simeq30.9$, where the breathing mode
frequency drops to zero. At the same time, the free energy $\mathcal{F}$
is always negative, indicating that the metastable droplet state found
at zero temperature does not exist anymore. The threshold number of
particle for the excitation-forbidden window also significantly decreases
and we find that $N_{\textrm{th}}(T=0.6T_{\textrm{th}})\simeq66.2$.
For $N>66.2$, the quadruple mode frequency $\omega_{20}$ decreases
with increasing number of particles, while the breathing mode frequency
$\omega_{00}$ continuously follows the particle-emission threshold
$\left|\mu\right|$ at the number of particles considered in the figure.

In Fig. \ref{fig3_CMLargeTemp}, we show the excitation spectrum at
an even larger temperature $T=0.8T_{\textrm{th}}$. At this temperature,
the excitation-forbidden window in the number of particles completely
disappears. Both the breathing mode frequency and quadruple mode frequency
appear to be bounded below the particle-emission threshold $\left|\mu\right|$.

\begin{figure}[t]
\begin{centering}
\includegraphics[width=0.48\textwidth]{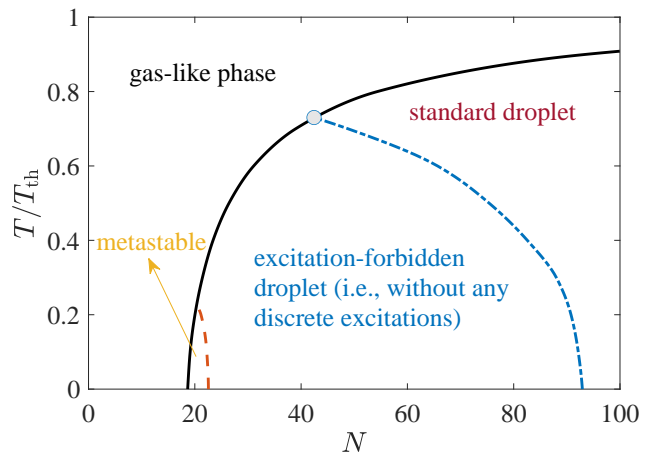} 
\par\end{centering}
\caption{\label{fig4_phasediagram} The phase diagram of a finite-size Bose
droplet as functions of the reduced number of particles $N$ (horizontal
axis) and of the reduced temperature $T/T_{\textrm{th}}$ (vertical
axis). At large temperature and small number of particles, the system
is in the gas-like phase, while at low temperature and large number
of particles, it is in the droplet state. A metastable droplet state
also occurs at low temperature and relatively small number of particle.}
\end{figure}

\subsection{A finite-temperature phase diagram }

We have calculated the excitation spectrum at different reduced temperatures
and consequently have obtained a finite-temperature phase diagram,
as reported in Fig. \ref{fig4_phasediagram}. This presents the main
result of our work. Here, the critical number of particle $N_{c}$
(black solid curve) is determined by extrapolating the breathing mode
frequency $\omega_{00}$ to zero, and the critical number $N_{m}$
(red dashed curve) is obtained by tracing the position where the free
energy becomes positive. The two curves crosses with each other at
about $0.24T_{\textrm{th}}$, above which the window for a metastable
droplet state closes. It is interesting to note that the critical
number of particles $N_{c}$ shows a sensitive temperature dependence.
It increases very rapidly once the temperature is above about $0.4T_{\textrm{th}}$.
Approaching the bulk threshold temperature $T_{\textrm{th}}$, a Bose
droplet with any number of particles becomes thermally unstable, as
we already show in the previous work \cite{Wang2020}.

On the other hand, the threshold number of particle $N_{\textrm{th}}$
(blue dash-dotted curve) can be determined from the crossing point
between the quadruple mode frequency $\omega_{20}$ and the particle-emission
threshold $\left|\mu\right|$. It separates the phase space for a
stable Bose droplet into two regimes: an\emph{ excitation-forbidden}
droplet regime without any bounded collective excitations below the
particle-emission threshold and a \emph{standard} droplet regime with
at least one discrete collective excitation. With increasing temperature,
we find that the $N_{\textrm{th}}$-curve terminates at about $0.73T_{\textrm{th}}$
(see, i.e., the solid circle symbol in the figure). Above this temperature,
we always find standard Bose droplets, in which a small but nonzero
temperature or entropy could be accommodated by the discrete bounded
collective excitations. Therefore, the intriguing self-evaporation
phenomenon predicted by Petrov, i.e., the emission of particles upon
arbitrary excitations \cite{Petrov2015}, ceases to exist. The Bose
droplets then fail to automatically reach zero temperature.

\begin{figure}[t]
\begin{centering}
\includegraphics[width=0.48\textwidth]{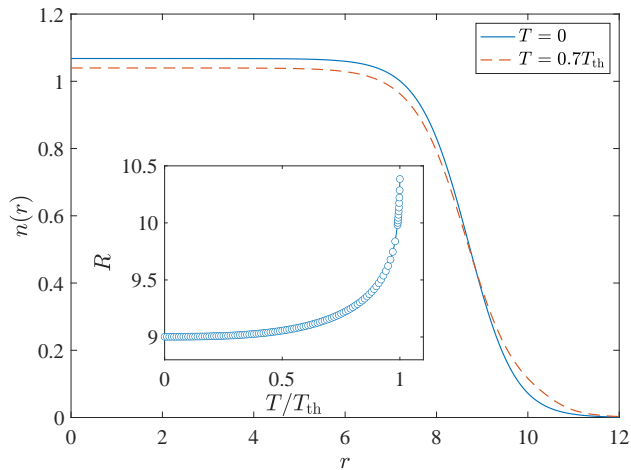} 
\par\end{centering}
\caption{\label{fig5_dstyNT3000} The density distribution $n(r)$ of a large
self-bound Bose droplet ($N=3000$) at zero temperature (blue solid
curve) and at the temperature $T=0.7T_{\textrm{th}}$ (red dashed
curve). The inset shows the temperature dependence of the size of
the Bose droplet, defined by $R=\sqrt{5\left\langle r^{2}\right\rangle /3}$.}
\end{figure}

\begin{figure}[t]
\begin{centering}
\includegraphics[width=0.48\textwidth]{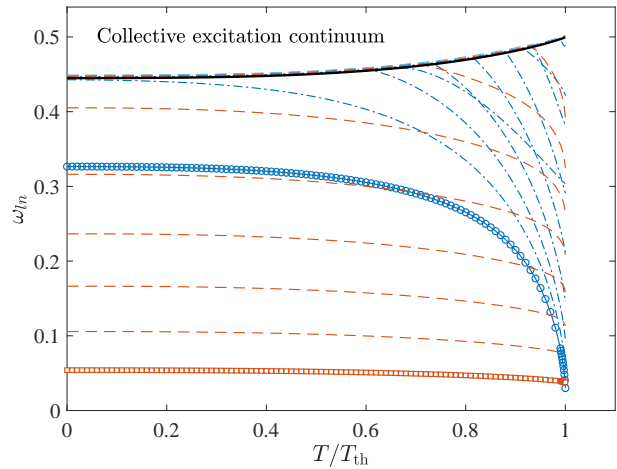} 
\par\end{centering}
\caption{\label{fig6_CMNT3000} Excitation frequencies $\omega_{ln}$ ($l\protect\leq9$
and $n\protect\leq2$) of a large self-bound Bose droplet ($N=3000$),
as a function of the reduced temperature $T/T_{\textrm{th}}$. The
red dashed curves show the surface modes $\omega_{l\protect\geq2,n=0}$
and the blue dash-dotted curves show the compressional bulk modes.
The lowest surface mode $\omega_{20}$ (i.e, quadruple mode) and the
lowest bulk mode (breathing mode) are highlighted by the red open
squares and blue open circles, respectively. The black thick curve
shows the threshold $-\mu$.}
\end{figure}

\subsection{The temperature-dependences of the density distribution and collective
excitations}

Let us now consider an \emph{idealized} experimental situation. Initially,
a Bose droplet is nearly in thermal equilibrium at a nonzero temperature.
It then gradually reduces its temperature by emitting a very small
portion of the most energetic particles. This slow self-evaporation
might be treated as an adiabatic process. By taking the in-situ or
time-of-flight absorption imaging of the Bose droplet, we may then
experimentally extract the temperature-dependences of the density
distribution and collective excitations of the Bose droplet, at a
nearly constant number of particles.

\subsubsection{Large Bose droplets}

In Fig. \ref{fig5_dstyNT3000}, we present the density distribution
of a large Bose droplet with the number of particles $N=3000$ at
zero temperature (solid curve) and at $T=0.7T_{\textrm{th}}$ (dashed
curve). For such a large droplet, the distribution acquires the typical
flat-top structure \cite{Petrov2015}. Moreover, the temperature dependence
of the density profile is not so apparent: the difference between
the distributions at the two temperatures is less than $3\%$. This
is correlated with a weak-temperature dependence of the droplet radius
$R$, as shown in the inset, where the radius is defined as the square
root of the mean square of the distance, $R=\sqrt{5\left\langle r^{2}\right\rangle /3}$.
The droplet size only increases notably when the temperature is close
to the bulk thermal destabilization threshold $T_{\textrm{th}}$ (i.e.,
$T>T_{\textrm{th}}$).

In Fig. \ref{fig6_CMNT3000}, we report the corresponding collective
excitation spectrum as a function of the temperature. At zero temperature,
there are a number of discrete modes, which can be well categorized
as the surface modes ($\omega_{l\geq2,n=0}$, red open squares and
red dashed curves) and bulk modes ($\omega_{00}$, the breathing mode
in blue circles; and $\omega_{l,n\neq0}$, blue dash-dotted curves)
\cite{Hu2020d}. The surface modes only propagate near the edge of
the Bose droplet and have an exotic dispersion relation, $\omega_{l0}^{(\textrm{surface})}\propto\sqrt{\sigma_{s}l(l-1)(l+2)}/R^{3/2}$,
with $\sigma_{s}$ being the surface tension \cite{Barranco2006,Petrov2015}.
In contrast, the bulk modes are compressional sound modes that propagate
through the whole droplet and have the standard dispersion relation,
$\omega_{ln}^{(\textrm{bulk})}\propto c/R$, where $c$ is the bulk
sound velocity \cite{Barranco2006,Hu2020d}. It is readily seen that
the frequency of the low-lying surface modes does not change too much
with increasing temperature. This could be understood from the robust
flat-top and temperature insensitive density distribution as we observe
in Fig. \ref{fig5_dstyNT3000}. As a result, the droplet size $R$,
the surface tension $\sigma_{s}$ and hence the surface mode frequencies
are less dependent on the temperature. On the other hand, the frequency
of the bulk sound modes has a strong temperature dependence and clearly
shows a waterfall-like effect close to the bulk threshold temperature
$T_{\textrm{th}}$. The is related to the the softening of the sound
velocity, which becomes exactly zero at the droplet-to-gas transition.
All the bulk mode frequencies therefore have to vanish towards the
transition.

\subsubsection{Small Bose droplets}

\begin{figure}[t]
\begin{centering}
\includegraphics[width=0.48\textwidth]{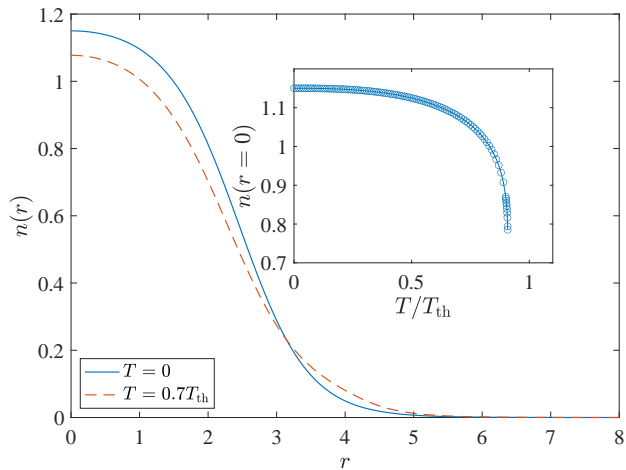} 
\par\end{centering}
\caption{\label{fig7_dstyNT100} The density distribution $n(r)$ of a small
self-bound Bose droplet ($N=100$) at zero temperature (blue solid
curve) and at the temperature $T=0.7T_{\textrm{th}}$ (red dashed
curve). The inset shows the temperature dependence of the center density
at $r=0$.}
\end{figure}

\begin{figure}[t]
\begin{centering}
\includegraphics[width=0.48\textwidth]{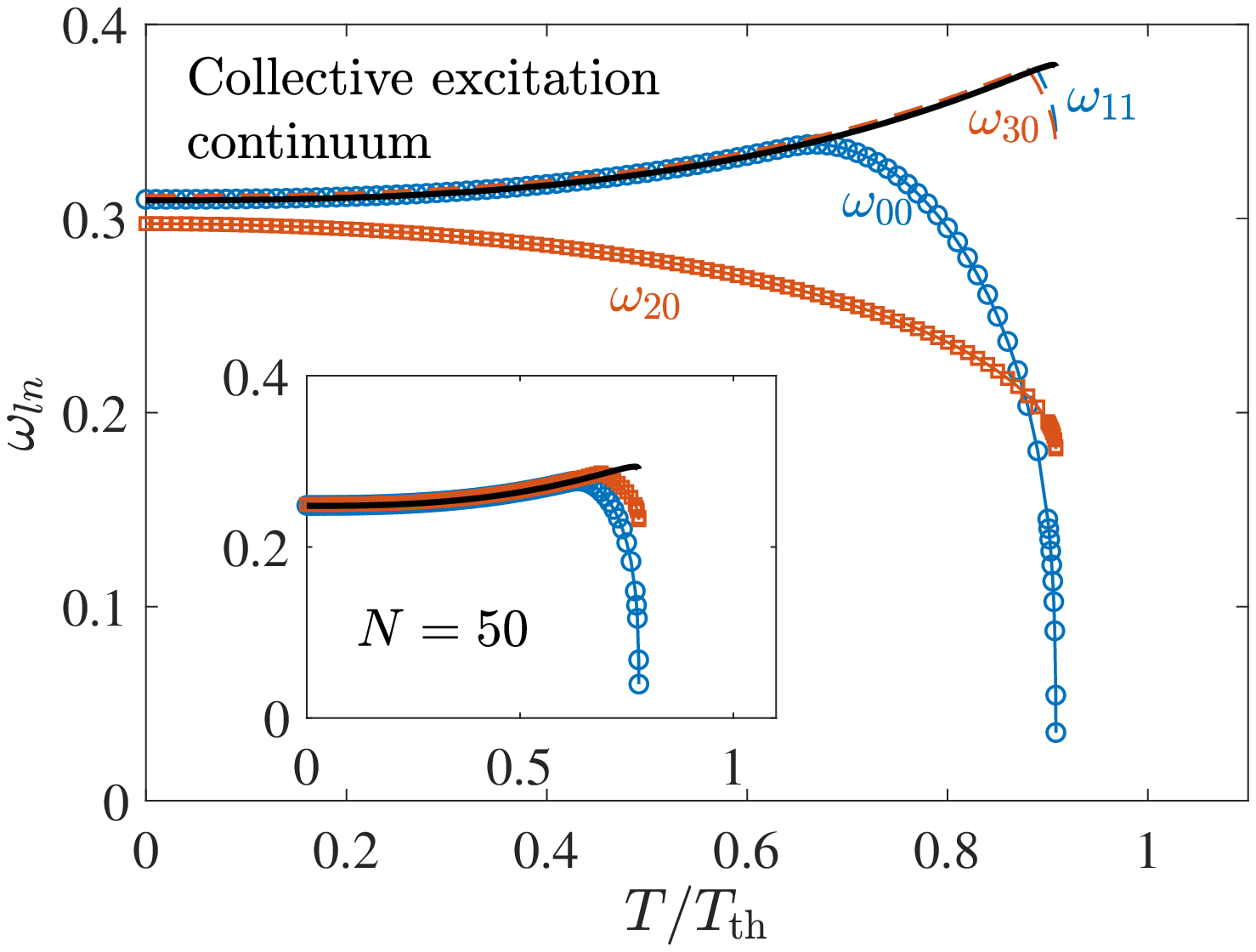} 
\par\end{centering}
\caption{\label{fig8_CMNT100} Excitation frequencies $\omega_{ln}$ ($l\protect\leq9$
and $n\protect\leq2$) of a small self-bound Bose droplet ($N=100$),
as a function of the reduced temperature $T/T_{\textrm{th}}$. The
red dashed curves show the surface modes $\omega_{l\protect\geq2,n=0}$
and the blue dash-dotted curves show the compressional bulk modes.
The lowest surface mode $\omega_{20}$ (i.e, quadruple mode) and the
lowest bulk mode (breathing mode) are highlighted by the red open
squares and blue open circles, respectively. The black thick curve
shows the threshold $-\mu$. At this number of particles, most of
the excitation modes enter the collective excitation continuum with
a frequency $\omega_{ln}\apprge-\mu$. The inset shows the excitation
spectrum at an even smaller number of particles, $N=50$.}
\end{figure}

Let us now consider a Bose droplet with small reduced number of particles,
which is more amenable to be created in the current experimental setups
\cite{Cabrera2018,Semeghini2018}. In Fig. \ref{fig7_dstyNT100},
we show the density distribution of a $N=100$ Bose droplet at two
temperatures: $T=0$ (solid curve) and $T=0.7T_{\textrm{th}}$ (dashed
curve). Compared with a large Bose droplet in Fig. \ref{fig5_dstyNT3000},
the density distribution of a small droplet shows a more appreciable
temperature dependence. In particular, the center density can change
up to several tens of percent (see the inset), as we increase the
temperature towards the threshold. This pronounced temperature dependence
could be related to the loss of the flat-top structure in the density
distribution due to the reduced number of particles. A small Bose
droplet appears to be more easier to be altered than a large droplet.

In Fig. \ref{fig8_CMNT100}, we report the temperature evolution of
the collective excitation spectrum at $N=100$. At this number of
particles and at zero temperature, only the lowest surface mode (i.e.,
the quadruple mode $\omega_{20}$) is bounded below the particle-emission
threshold $\left|\mu\right|$ \cite{Petrov2015}. When we increase
temperature, the quadruple mode frequency decreases notably, presumably
due to the increase of the droplet radius, since the droplet at this
size acquires a more sensitive temperature dependence as we mentioned
earlier. Interestingly, at about $0.7T_{\textrm{th}}$ the frequency
of the lowest compression bulk mode, the breathing mode frequency,
starts to fall off the particle-emission threshold. It becomes increasingly
softened towards the threshold temperature $T_{\textrm{th}}$. At
an even higher temperature (i.e., $T\sim0.87T_{\textrm{th}}$), more
and more higher-order bulk modes fall off the the particle-emission
threshold and becomes softened. This fall-off feature turns out to
be very general, occurring also at smaller number of particles, as
can be seen in the inset for the selected case of $N=50$.

The mode frequency softening, for both surface modes and compressional
bulk modes in large and small Bose droplets, is therefore a characteristic
feature of the thermally-induced droplet-to-gas transition at finite
temperature. The mode softening effectively removes the excitation-forbidden
interval in the number of particles predicted by Petrov at zero temperature
\cite{Petrov2015}, and opens the possibility to observe a small Bose
droplet with non-zero temperature.

\section{Conclusions}

In summary, we have theoretically investigated the finite-temperature
effects on the structure and collective excitations of an ultradilute
quantum droplet in free space, formed in a binary Bose-Bose mixture
with inter-species attractions near the mean-field collapse. Our calculations
are based on the extended (time-dependent) Gross-Pitaevskii equation
generalized to the finite-temperature case. The density distribution
is determined by solving the static Gross-Pitaevskii equation, while
the collective excitation spectrum is obtained by solving the coupled
Bogoliubov equations.

We have found a rich finite-temperature phase diagram as a function
of the number of particles in the droplet. In particular, the critical
number of particles at the droplet-to-gas transition is found to depend
sensitively on the temperature. The excitation-forbidden interval
predicted by Petrov is shown to shrink with increasing temperature
and disappear completely at about 0.73$T_{\textrm{th}}$, where $T_{\textrm{th}}$
is the threshold temperature for thermally destabilizing an infinitely
large Bose droplet. Above the temperature 0.73$T_{\textrm{th}}$,
there is at least one discrete collective mode below the particle-emission
threshold, which may block the self-evaporation of the Bose droplet
and allow a small but nonzero temperature.

Our results could be experimentally examined, if we are able to overcome
the difficulty of finding a useful thermometry to measure the temperature.
Qualitatively, at the number of particles slightly below $N_{\textrm{th}}(T=0)\simeq94.2$,
the experimental observation of discrete quadrupole mode frequency
or breathing mode frequency below the particle-emission threshold,
i.e., $\omega_{20}<\left|\mu\right|$ or $\omega_{00}<\left|\mu\right|$,
would be a very strong evidence for the finite-temperature effect. 
\begin{acknowledgments}
This research was supported by the Australian Research Council's (ARC)
Discovery Program, Grants No. DE180100592 and No. DP190100815 (J.W.),
Grant No. DP180102018 (X.-J.L), and Grant No. DP170104008 (H.H.). 
\end{acknowledgments}

\appendix

\section{Numerical method}

Here, we describe our numerical approach to find the ground state
solution of the static extended GPE Eq. (\ref{eq:stationaryEGPE}),
which is equivalent to minimize an energy density functional $\mathcal{E}/N$,
where $N=\int d{\mathbf{r}}|\phi_{0}(\mathbf{r})|^{2}$. The energy
functional in the free-space is given by, 
\begin{equation}
\mathcal{E}=\langle K\rangle+\langle\mathcal{F}\rangle,
\end{equation}
where 
\begin{equation}
\langle K\rangle=-\frac{1}{2}\int d{\mathbf{r}}\phi_{0}(\mathbf{r})\nabla^{2}\phi_{0}(\mathbf{r}),
\end{equation}
and 
\begin{equation}
\langle\mathcal{F}\rangle=\int d{\mathbf{r}}\mathcal{F}(\mathbf{r}).
\end{equation}
Here $\mathcal{F}(\mathbf{r})$ is given by Eq. (\ref{eq:DimensionlessFreeEnergy}),
with $\mathbf{r}$-dependency explicitly written out.

Due to the spherical symmetry of the system, we only consider the
$s$-wave solution $(l=0)$. Therefore, Eq.(\ref{eq:stationaryEGPE})
reduces to an effective 1D radial equation. We first expand the condensed
wave function in a $5$-th order $B$-splines basis \cite{Boor1978}:
\begin{equation}
\phi_{0}(\mathbf{r})=\frac{1}{\sqrt{4\pi}}\sum_{n}c_{n}\frac{b_{n}(r)}{r}.
\end{equation}
$B$-spline basis has been extensively used in solving Schr{ö}dinger
equations in two- and three-body problems with high accuracy \cite{Hart1997,Wang2010}.
$B$-spline basis allows to use an uneven grid, which might better
represent the solution wave-function. $B$-spline basis also allows
a higher order approximation of the derivative operator that appears
in the kinetic energy term. The energy density functional $\mathcal{E}/N$
then can be regarded as a non-linear function of coefficients $c_{n}$,
which can be minimized using the standard conjugate-gradient method
via software package such as ``minFunc'' in Matlab \cite{Schmidt2005}.

\end{document}